Frequency decoding of periodically timed action potentials
through distinct activity patterns in a random neural network

*Short title:* Frequency decoding in a random neural network

Tobias Reichenbach and A. J. Hudspeth*

Howard Hughes Medical Institute and Laboratory of Sensory Neuroscience,
The Rockefeller University, New York, NY 10065, USA

*\* To whom correspondence should be addressed; E-mail: hudspaj@rockefeller.edu.*

## Abstract

**Frequency discrimination is a fundamental task of the auditory system. The mammalian inner ear, or cochlea, provides a place code in which different frequencies are detected at different spatial locations. However, a temporal code based on spike timing is also available: action potentials evoked in an auditory-nerve fiber by a low-frequency tone occur at a preferred phase of the stimulus—they exhibit phase locking—and thus provide temporal information about the tone's frequency. In an accompanying psychoacoustic study, and in agreement with previous experiments, we show that humans employ this temporal information for discrimination of low frequencies. How might such temporal information be read out in the brain? Here we demonstrate that recurrent random neural networks in which connections between neurons introduce characteristic time delays, and in which neurons require temporally coinciding inputs for spike initiation, can perform sharp frequency discrimination when stimulated with phase-locked inputs. Although the frequency resolution achieved by such networks is limited by the noise in phase locking, the resolution for realistic values reaches the tiny frequency difference of 0.2% that has been measured in humans.**

# Author Summary

Humans can resolve tiny frequency differences of only 0.2%, much below the frequency interval of a semitone in Western music which is about 6%. How is this astonishing frequency resolution achieved?

Sound is detected within the inner ear, or cochlea, in which auditory-nerve fibers fire action potentials upon acoustic stimulation. Which of an ear's 30,000 auditory-nerve fibers fire indicates the frequency of a pure tone. However, the timing of spikes from a single auditory-nerve fiber can also provide information about the signal's frequency. Recent psychoacoustic experiments on the human perception of tones show that humans indeed employ this temporal information. The neural mechanisms, however, remain unclear.

Here we show that a class of neural networks—with random connections, temporal delays, and coincidence detection—can read out the frequency information provided in the spike timing of auditory neurons. We employ methods from statistical physics as well as numerical simulations to demonstrate the frequency resolution that such networks can achieve resembles that of human observers.

# Introduction

A sound impinging on the eardrum elicits a wave of displacement of the basilar membrane within the cochlea [1,2]. Mechanosensitive hair cells on the basilar membrane transduce the membrane's vibration into electrical signals that are transmitted to the associated auditory-nerve fibers [3,4]. Through position-dependent resonance along the basilar membrane the cochlea establishes a place code for frequencies: high frequencies evoke traveling waves that peak near the organ's base whereas the waves elicited by lower frequencies culminate at more apical positions.

A temporal code may, however, supplement or even supersede the place code. In response to a pure sound at a frequency below about 300 Hz, an auditory-nerve fiber fires action potentials at every cycle of stimulation and at a fixed phase [3,5]. Above 300 Hz the axon starts to skip cycles, but action potentials still occur at a preferred phase of the stimulus. The quality of this phase locking decays between 1 kHz and 4 kHz, however, and phase locking is lost for still higher frequencies. Phase locking below 4 kHz is sharpened in the auditory brainstem by specialized neurons such as spherical bushy cells that receive input from multiple auditory-nerve fibers [6,7]. These cells can fire action potentials at every cycle of stimulation up to 800 Hz (Supplementary Figure 1). Temporal information about the stimulus frequency is therefore greatest for frequencies below 800 Hz, declines from 800 Hz to 4 kHz, and vanishes for still greater frequencies. In some species, such as the barn owl, phase locking can continue up to 10 kHz [8].

Phase locking is employed for sound localization in the horizontal plane [9,10]. A sound coming from a subject's left, for example, reaches the left ear first and hence produces a phase delay in the stimulus at the right ear compared to that at the left. Auditory-nerve fibers preserve this phase difference, which is subsequently read out by binaurally sensitive neurons through delay and coincidence detection. A temporal delay generally results when one neuron signals another: the signal propagates along the axon of the transmitting neuron and along the dendrites of the receiving cell, producing delays of up to 20 ms with only a few microseconds of jitter [11-13]. Coincidence detection occurs when two or more synchronous incoming spikes are required for a neuron to fire: the signals must arrive at the nerve cell's soma within a certain time window $\tau$, comparable to the membrane's time constant, in order for their effects to add and initiate an action potential.

Phase locking can also provide information about the frequency of a pure tone, for the duration between two subsequent neural spikes is on average the signal's period or a multiple thereof. Evidence for the usage of this information in the brain comes from psychoaoustic studies that show that human frequency discrimination is superior for the lower frequencies at which phase locking is available and that discrimination of these frequencies worsens when the phase information is perturbed [14-18].

It remains unclear how the temporal information on frequency is read out in the brain. The usage of delay and coincidence detection has been proposed, but the exact mechanism has not been specified and the resulting frequency resolution has not been determined [19]. Other studies have proposed mathematical schemes for determining a signal's frequency from phase locking, the neural implementation of which remains unclear [20,21]. Here we study how a random recurrent neural network with delay and coincidence detection can encode frequency in its activity pattern when stimulated with phase-locked, cycle-by-cycle input.

## Results

Denote by $T$ the period and by $L$ the duration in cycles of a signal such that the phase-locked, cycle-by-cycle external spikes arrive at the network neurons around times $0, T, 2T, 3T, ..., (L\text{-}1)T$. Assume that the first external spike triggers an action potential at each neuron; because of adaption, two coincident spikes are needed for the generation of subsequent spikes. If a neuron $i$ projects to another neuron $j$ then its first spike arrives there at a later time $t_{ij}$ that represents the delay between the cells. If that time differs no more than the small amount $\tau$ from the time $T$ at which the second external spike arrives at neuron $j$, the two spikes act in concert to elicit an action potential; otherwise neuron $j$ remains silent. If active, neuron $j$ may trigger spikes in other neurons, specifically those for which the time delay from neuron $j$ also matches the signal period. Sustained network activity results when the connectivity $C$ between neurons—the average number of internal connections that a neuron receives—exceeds a certain value (Figure 2; Supplementary Methods).

How can we quantify this network's pattern of activity? Let the network comprise $N$ neurons and denote a neuron as active if it fires spikes in response to at least half of the external spikes and as inactive otherwise. The network's activity may then be summarized by a binary vector $\mathbf{x}_T = \left( x_T^{(1)}, x_T^{(2)}, ..., x_T^{(N)} \right)$ in which $x_T^{(i)} = 1$ if neuron $i$ ($i$=1, 2, ..., $N$) is active under stimulation at a period $T$ and $x_T^{(i)} = 0$ otherwise. The fraction $a$ of active nodes follows as

$$a = \frac{1}{N} \sum_{i=1}^{N} x_T^{(i)}. \tag{1}$$

An analytical approximation provides insight into the dependence of the network's activity on its connectivity and size. Assume that a neuron $j$ is active if it receives at least one active connection, in which we define a connection from another neuron $i$ to neuron $j$ as active if spikes from neuron $i$ traveling to neuron $j$ can elicit action potentials there in at least half of the trials. Denote the average number of active connections that a neuron receives by $B$. The probability that a neuron does not receive any active connection then reads $\left[ 1 - B/(N-1) \right]^{a(N-1)}$ and equals the fraction of inactive neurons:

$$1 - a = \left[ 1 - B/(N-1) \right]^{a(N-1)} \approx e^{-aB}. \tag{2}$$

The approximation can be solved through the Lambertz W-function,

$$a = 1 + \frac{1}{B} W \left( -Be^{-B} \right). \tag{3}$$

Further analysis shows that the average number of active connections can be approximated as $B = 2\tau C / (t_{max} - t_{min})$ and the analytically derived average network activity is then in excellent agreement with numerical simulations (Figure 1c and Supplementary Methods). The fraction of active nodes does not depend on the network size $N$ if the network connectivity $C$ is independent of $N$. Because the probability $c$ of a connection between two neurons follows as $c = C/(N-1)$ the resulting networks are sparse. We denote the network connectivity at which half of the neurons are active by $C_*$. Because an activity pattern is most informative when half of the neurons are active, we employ this connectivity in the following.

Inaccuracy in phase locking results in noise: spikes from spherical bushy cells, for example, exhibit a phase distribution that is approximately Gaussian around the mean value with a standard deviation $s$ that can be as small as one-twentieth of a cycle [7]. Input spikes therefore arrive at the network neurons at times $\xi_1$, $T+\xi_2$, ..., $(L-1)T+\xi_L$ in which $\xi_k$ ($k = 1, 2, ..., L$) is a random Gaussian variable with zero mean and standard deviation $s$. Although small, this noise evokes slightly different neural activity patterns upon repeated stimulation. The brain may nevertheless learn the mean pattern at period $T$ over many repeated trials. We define this mean pattern as $\boldsymbol{X}_T = \left(X_T^{(1)}, X_T^{(2)}, ..., X_T^{(N)}\right)$ in which $X_T^{(i)} = 1$ if neuron $i$ ($i = 1, 2, ..., N$) is active in at least half of the trials and $X_T^{(i)} = 0$ otherwise.

Stimulation at another signal period $T'$ evokes a different mean activity pattern $\boldsymbol{X}_{T'}$. We can quantify its difference from the mean pattern $\boldsymbol{X}_T$ at period $T$ through the relative Hamming distance $d(\boldsymbol{X}_T, \boldsymbol{X}_{T'})$ between the patterns:

$$d(\boldsymbol{X}_T, \boldsymbol{X}_{T'}) = \frac{1}{N} \sum_{i=1}^{N} \left| X_T^{(i)} - X_{T'}^{(i)} \right|. \tag{4}$$

This distance specifies the fraction of neurons that differ in their activity between the two patterns. Analytical calculations and numerical simulations show that the distance increases linearly in the absolute period difference $|T'-T|$ and vanishes for $T' \doteq T$ (Figure 3a; Supplementary Methods). The network thus decodes stimulation periods through distinct patterns of mean activity. Each such pattern may then selectively activate a particular downstream neuron.

The identification of the period from an individual signal is inevitably limited by the noise in the timing of the external spikes. When a network is stimulated at a period $T$ its single-trial activity pattern $\boldsymbol{x}_T$ differs both from the mean pattern $\boldsymbol{X}_T$ at period $T$ and from the mean pattern $\boldsymbol{X}_{T'}$ at another period $T'$. Correct discrimination between $T$ and $T'$ thus requires the pattern $\boldsymbol{x}_T$ to be closer to $\boldsymbol{X}_T$ then to $\boldsymbol{X}_{T'}$.

The relative Hamming distance $d(\boldsymbol{x}_T, \boldsymbol{X}_T)$ between $\boldsymbol{x}_T$ and $\boldsymbol{X}_T$ is the average over the $N$ random variables $\left| x_T^{(i)} - X_T^{(i)} \right|$. An analytical approximation shows that these variables can be regarded as effectively independent. The distribution of $d(\boldsymbol{x}_T, \boldsymbol{X}_T)$ is therefore Gaussian around a mean value $D(\boldsymbol{x}_T, \boldsymbol{X}_T)$ with a certain standard deviation $\sigma$ (Figure 3b; Supplementary Methods). The distance $d(\boldsymbol{x}_T, \boldsymbol{X}_{T'})$ between $\boldsymbol{x}_T$ and $\boldsymbol{X}_{T'}$ is also Gaussian around another mean value $D(\boldsymbol{x}_T, \boldsymbol{X}_{T'})$ but with the same standard deviation $\sigma$ as for $d(\boldsymbol{x}_T, \boldsymbol{X}_T)$ (Figure 3d). The two distributions can be differentiated with at least 95% accuracy when the mean values $D(\boldsymbol{x}_T, \boldsymbol{X}_T)$ and $D(\boldsymbol{x}_T, \boldsymbol{X}_{T'})$ differ by $4\sigma$ or more. When is this condition fulfilled?

Because larger system sizes $N$ imply averaging over more neurons, and as confirmed by analytical and numerical computations, the standard deviation $\sigma$ decreases as $N^{-1/2}$ in accordance with the central limit theorem (Figure 3c; Supplementary Methods). We can therefore resort to a network that is large enough to yield a sufficiently small variance in the relative Hamming distance. Correct discrimination of a signal's period between $T$ and $T'$ is then feasible as soon as the mean value $D(\boldsymbol{x}_T, \boldsymbol{X}_T)$ is distinct from $D(\boldsymbol{x}_T, \boldsymbol{X}_{T'})$.

How does the mean value $D(\boldsymbol{x}_T, \boldsymbol{X}_{T'})$ depend on the difference in periods? Analytical and numerical computations show that $D(\boldsymbol{x}_T, \boldsymbol{X}_{T'})$ increases linearly in $|T'\text{-}T|$ when $|T'\text{-}T| > \Delta T$ for a threshold difference $\Delta T$ (Figure 3d; Supplementary Methods). When the periods of the two signals are closer, $|T'\text{-}T| < \Delta T$, the distance $D(\boldsymbol{x}_T, \boldsymbol{X}_{T'})$ is proportional to the squared period difference $(T'-T)^2$ and approaches $D(\boldsymbol{x}_T, \boldsymbol{X}_T)$ for $T'=T$. The threshold $\Delta T$ thus provides a measure for the smallest period difference that a particular network can resolve.

The threshold value $\Delta T$ is proportional to the mean distance $D(\boldsymbol{x}_T, \boldsymbol{X}_T)$ between a single-trial pattern $\boldsymbol{x}_T$ and the mean pattern $\boldsymbol{X}_T$ at period $T$ (Figure 4; Supplementary Methods). Because the pattern evoked by a signal of greater length $L$ involves more averages over external spike times, and hence over the phase noise, it results in a smaller mean value $D(\boldsymbol{x}_T, \boldsymbol{X}_T)$ that decreases as $L^{-1/2}$ (Figure 5a; Supplementary Methods). However, a larger network size $N$ does not affect $D(\mathbf{x}_T, \mathbf{X}_T)$ (Figure 3c).

The threshold $\Delta T$ also decreases as $L^{-1/2}$ with increasing signal length $L$ (Figure 5b). Longer signals indeed provide more information that can enhance frequency resolution. The improvement in resolution is less than that expected from the Fourier uncertainty principle, in which frequency resolution is inversely proportional to signal length [22].

The threshold $\Delta T$ for a given signal length is proportional to the noise in the phase-locked input signal but independent of the neural network's parameters (Supplementary Methods). For a realistic value for the standard deviation $s$ of one-twentieth of a cycle and for a signal of length $L$=200 cycles, we obtain a period resolution $\Delta T/T$ of about 0.2% (Figure 5b). This value agrees well with the human frequency resolution measured in psychoacoustic experiments [15].

## Discussion

Our results demonstrate that the timing of action potentials, even without the cochlear place code, allows for accurate frequency discrimination. A simple, randomly connected network with a range of signal-propagation delays between neurons can use phase-locked inputs to replicate the striking frequency discrimination of the human auditory system.

The network encodes different input frequencies in distinct activity patterns. We have defined those patterns in the simplest possible way, through the mean activity of the network neurons during stimulation. Further statistics of the spike trains fired by the network neurons, such as temporal correlation between the spikes from one neuron as well as correlation between spikes from different neurons, can lead to finer discrimination of activity patterns and hence improve the precision in frequency discrimination.

The activity patterns that we have defined here may be read out by downstream neurons. A given downstream neuron may detect a particular activity pattern of the network if it receives excitatory connections from the neurons that are active for this pattern and inhibitory connections from the network neurons that are inactive for this pattern. The properties and precision of such a downstream read-out will be investigated in future studies.

In our simulations we have employed a range of temporal delays between network neurons that encompasses about an octave. Frequency discrimination by such a network is accordingly restricted to a spectral band of less than an octave, and many networks, each with a distinct range of temporal delays, are required to cover a broader frequency range. Where might such structures exist in the brain? The inferior colliculus displays a tonotopic array of multiple frequency-band laminae, each of which analyzes about one-third of an octave [23,24]. Substantial signal processing appears to be performed within each lamina, potentially including pitch detection [25,26]. Frequency discrimination through frequency-dependent network activity patterns as proposed here might therefore occur in these laminae. Simultaneous recordings from many interconnected neurons within one lamina would be required for an experimental test of this hypothesis.

Neural networks can exhibit emergent computational abilities such as synchronous information transmission, memory, and speech recognition that are not present at the level of individual nerve cells [27-32]. It remains uncertain to what extent such brain functions depend upon a neuron's precise action-potential timing as opposed to its average firing rate [33,34]. Although our study is specific to the auditory system, it may also help a more general understanding of the usage of temporal codes for other brain functions.

## Methods

Random neuronal networks are constructed by assigning to every pair $(i, j)$, $i{\neq}j$, of neurons a connection from $i$ to $j$ with a low probability $c$. The average number of connections emerging from an individual neuron accordingly reads $C{=}c(N{-}1)$ and equals the average number of a neuron's incoming connections. To each connection from a neuron $i$ to another neuron $j$ we assign a time delay $t_{ij}$ that is drawn randomly between a minimal time $t_{\min}$ and a maximal time $t_{max}$, $t_{ij} \in [t_{min}, t_{max}]$. In our simulations we have employed $t_{min}$=1.2 ms and $t_{max}$=2.8 ms and a period $T$=2 ms.

The probability distributions in Figure 3b show a typical result from one random network. All other numerical results, including mean activity patterns and the statistics of distances between individual trials and mean patterns, have been obtained by averaging over at least 100 different random networks.

We measure the arrival of an action potential at a neuron's soma by the time at which the maximum of the depolarization occurs. Each neuron fires an action potential upon arrival of a signal's first external spike. The initiation of subsequent action potentials requires that two action potentials arrive at the neuron's soma within a time window $\tau$; in our simulations we have employed $\tau$=0.6 ms. Such a difference between generation of the first spike and later ones could result, for example, from adaptation in a neuron. Generation of an action potential is followed by a refractory period for the duration of which we have assumed 1.2 ms.

For simulations of network dynamics we have developed a fast, event-based algorithm that stores the propagating spikes and their arrival times at each neuron. At each step in the algorithm we then compute the earliest subsequent time at which a neuron fires a spike, determine to which neurons that spike propagates as well as the associated arrival times, and appropriately update the list of incoming spikes at those neurons. The mean activity patterns as well as the statistics of the pattern distance of a single trial from a mean pattern have been computed from at least 100 trials for each network realization.

## Acknowledgments

We thank L. Abbott for discussion as well as P. Kumar and D. Ó. Maoiléidigh for helpful comments on the manuscript. T. R. holds a Career Award at the Scientific Interface from the Burroughs Wellcome Fund; A. J. H. is an Investigator of Howard Hughes Medical Institute.

# Supplementary Information

The Supplementary Information includes Supplementary Methods and two Supplementary Figures.

# Figure Legends

**Figure 1. Schematic diagrams of a neural network with delay and coincidence detection.**
**a**, Each neuron (light gray) can receive phase-locked inputs from a preceding neuron through external nerve fibers (orange). The network neurons are randomly connected (light blue) with the indicated characteristic delays in signal propagation. To fire an action potential, a neuron requires two temporally coincident spikes, such as one from an external and one from an internal source. **b**, When a periodic signal arrives through the external nerve fibers (red), the internal connections whose signal delay is approximately matched to the signal period induce spikes in their target neurons, resulting in a pattern of active internal connections (dark blue) and active neurons (black borders). **c**, A different signal period evokes a distinct pattern of active connections and active neurons.

**Figure 2: Patterns of network activity. a**, Each horizontal line depicts the activity of a single neuron in the network. Because every cell fires a spike upon receiving the first input signal, the raster of action potentials displays a vertical black line at its outset. The generation of spikes subsequently requires two incoming action potentials that temporally coincide. The connections between neurons induce different delays, so an activity pattern results in which some neurons fire at almost every cycle whereas others remain silent. Noise in the timing of the external spikes introduces variation in the firing of each neuron. **b**, The fraction of active neurons depends on the mean connectivity $C$, the average number of internal connections that each neuron receives. An analytical approximation (black line) confirms that the fraction of active neurons is independent of the network size $N$. Half of the neurons are active at a connectivity $C_* \approx 1.85$.

**Figure 3: Distances of activity patterns evoked by different signal periods. a**, The average distance $d(\boldsymbol{X}_T, \boldsymbol{X}_{T'})$ between the mean activity pattern for period $T$ and that for period $T'$ increases

linearly in the absolute value $|T'-T|$ of the period difference. Here we have employed $N$=1,000 and $L$=100. **b**, The distributions of the distances $d(\boldsymbol{x}_T,\boldsymbol{X}_T)$ (black squares) and $d(\boldsymbol{x}_T,\boldsymbol{X}_{T'})$ (red circles) are approximately Gaussian around the mean values $D(\boldsymbol{x}_T,\boldsymbol{X}_T)$ (dashed black line) and $D(\boldsymbol{x}_T,\boldsymbol{X}_{T'})$ (dashed red line). This result has been obtained from a single network with $N$=1,000 neurons, a signal of $L$=10 cycles, and $(T-T')/T$=3%. Averaged over multiple networks, the standard deviation $\sigma$ (shading) is equal for the two configurations. **c**, The mean value $D(\boldsymbol{x}_T,\boldsymbol{X}_T)$ (black squares) is independent of the network size $N$, whereas the standard deviation $\sigma$ (gray circles) decreases as $N^{-1/2}$ for larger $N$. The results were obtained for a signal of length $L$=50 cycles. **d**, The dependence of the distribution of $d(\boldsymbol{x}_T,\boldsymbol{X}_{T'})$ on the period difference $|T'-T|$ is plotted for $N$=300 and $L$=20. The mean value $D(\boldsymbol{x}_T,\boldsymbol{X}_{T'})$ (red circles) has a minimum for $T'=T$, at which it reaches $D(\boldsymbol{x}_T,\boldsymbol{X}_T)$ (horizontal black line). For $|T'-T|$ above a threshold value $\Delta T$ the mean value $D(\boldsymbol{x}_T,\boldsymbol{X}_{T'})$ increases linearly in $|T'-T|$ (dashed blue line), whereas it exhibits a quadratic dependence below (blue shading). The standard deviation $\sigma$ of $d(\boldsymbol{x}_T,\boldsymbol{X}_{T'})$ (red shading) does not depend on the period difference and equals the standard deviation of $d(\boldsymbol{x}_T,\boldsymbol{X}_T)$ (gray shading).

**Figure 4: Influence of network size $N$ and signal length $L$.** The mean distance $D(\boldsymbol{x}_T,\boldsymbol{X}_{T'})$ (red circles) does not change with size $N$ but the standard deviation (red shading) is progressively reduced. Longer signals, however, induce a lower mean value $D(\boldsymbol{x}_T,\boldsymbol{X}_T)$ (black line) and hence a smaller value $\Delta T$ at which the crossover occurs from linear to quadratic dependence of $D(\boldsymbol{x}_T,\boldsymbol{X}_{T'})$ on $|T'-T|$ (blue shading). The standard deviations of $d(\boldsymbol{x}_T,\boldsymbol{X}_{T'})$ (red shading) and $d(\boldsymbol{x}_T,\boldsymbol{X}_T)$ (gray shading) also decline for longer signals.

**Figure 5: Dependence of period resolution on signal length $L$**. **a**, The mean value $D(\boldsymbol{x}_T,\boldsymbol{X}_T)$ (black squares) decreases as $L^{-1/2}$ with increasing signal length whereas the standard deviation $\sigma$ (gray circles) exhibits the weaker dependence $L^{-1/4}$. The results were computed for $N$=1,000. **b**, The period resolution $\Delta T$ achieved by a network with $N$=500 also decreases as $L^{-1/2}$.

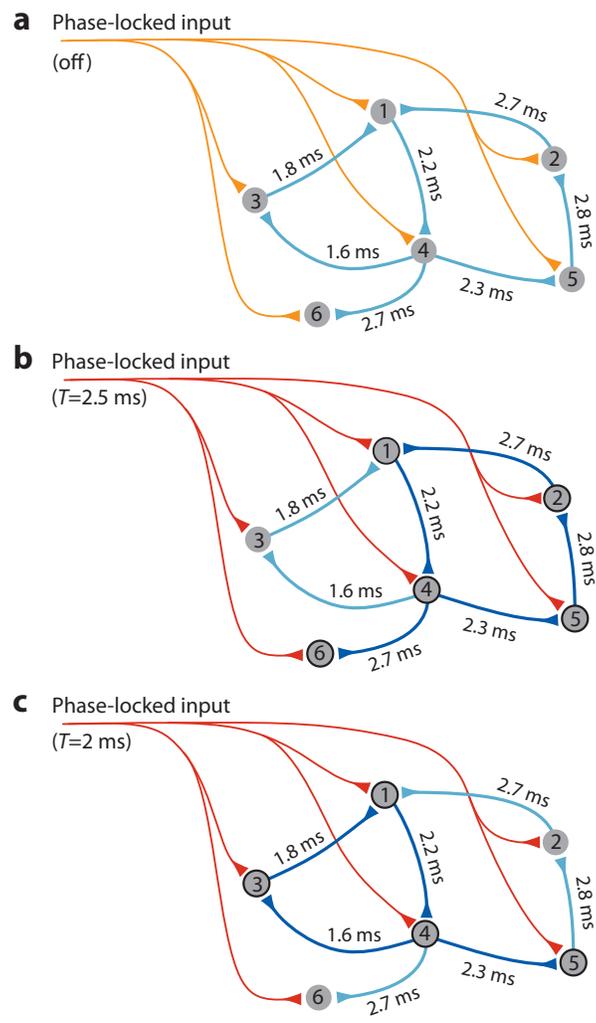

**a** Phase-locked input
(off)

**b** Phase-locked input
(*T*=2.5 ms)

**c** Phase-locked input
(*T*=2 ms)

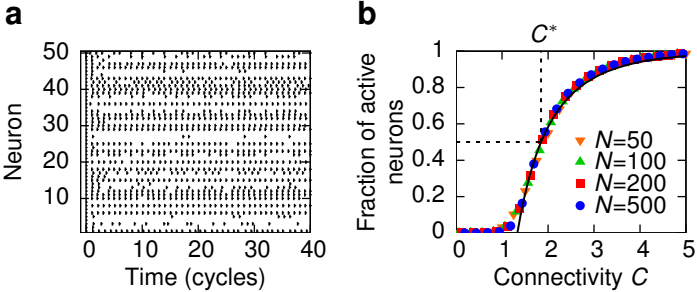

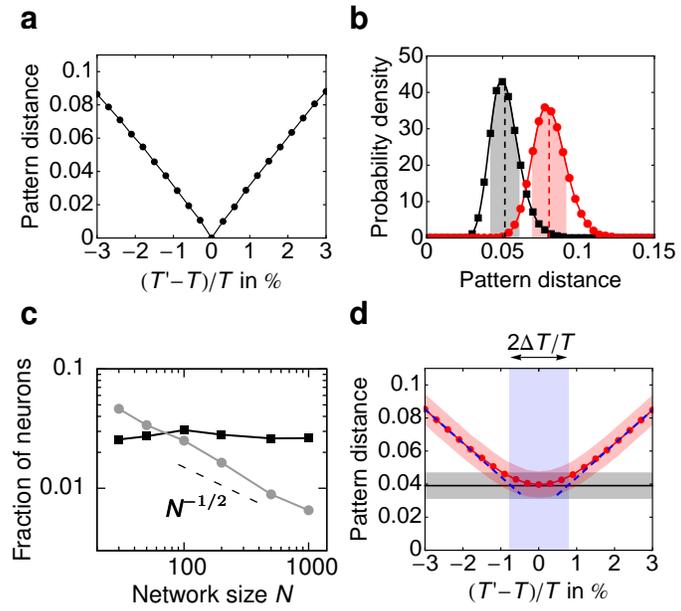

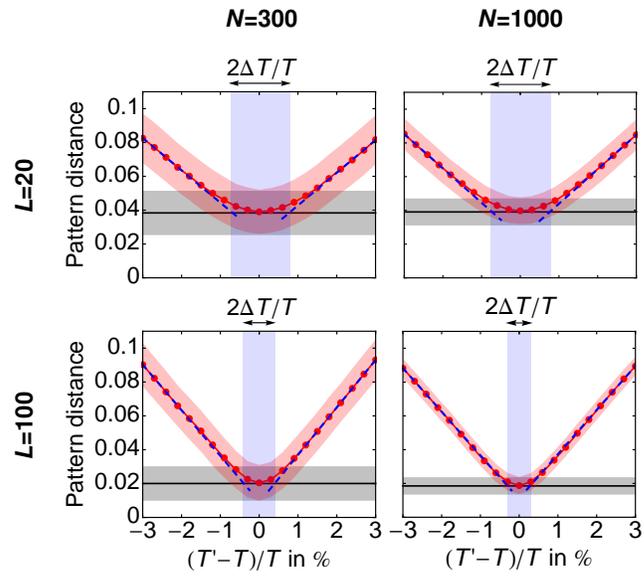

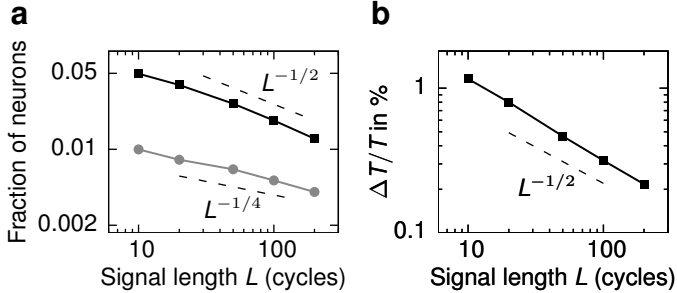

Supplementary Information:

Frequency decoding of periodically timed action potentials

through distinct activity patterns in a random neural network

Tobias Reichenbach and A. J. Hudspeth*

Howard Hughes Medical Institute and Laboratory of Sensory Neuroscience,

The Rockefeller University, New York, NY 10065, USA

*  *To whom correspondence should be addressed; E-mail: hudspaj@rockefeller.edu.*

## Supplementary Methods

### Analytical computation of the mean network activity

What is a neuron's average number $B$ of active connections? Suppose that neuron $i$ has an outward connection to excite neuron $j$ and the spike's travel time is $t_{ij}$. Assume that neuron $i$ fires at a time $nT + \xi_n$, such that the signal arrives at neuron $j$ at the time $nT + \xi_n + t_{ij}$. For spike initiation at neuron $j$ this time may differ by no more than a time $\tau$ from the arrival time $(n+1)T + \xi_{n+1}$ of the next external spike there. In other words,

$$\xi_n + t_{ij} \in [T + \xi_{n+1} - \tau, T + \xi_{n+1} + \tau] \tag{S1}$$

or

$$\xi_n - \xi_{n+1} \in [T - t_{ij} - \tau, T - t_{ij} + \tau]. \tag{S2}$$

Because $\xi_n$-$\xi_{n+1}$ is a stochastic process with zero mean and standard deviation $\sqrt{2}s$, the probability $p(T, t_{ij})$ that Equation S2 is fulfilled is

$$p(T, t_{ij}) = \frac{1}{2} \left[ \mathrm{erf}\left( \frac{T - t_{ij} + \tau}{2s} \right) - \mathrm{erf}\left( \frac{T - t_{ij} - \tau}{2s} \right) \right]. \tag{S3}$$

An external spike at neuron $j$ together with an internal spike received from neuron $i$ therefore causes neuron $j$ to fire with a probability $p(T, t_{ij})$. The connection from $i$ to $j$ is active if the number of such spikes associated with $L$ external spikes is at least $L/2$, and the corresponding probability $q(T, t_{ij})$ is

$$q(T, t_{ij}) = \frac{1}{2} \left[ 1 + \mathrm{erf}\left( \frac{p(T, t_{ij}) - L/2}{\sqrt{2p(T, t_{ij})[1 - p(T, t_{ij})]/L}} \right) \right]. \tag{S4}$$

Denote the index set of neurons that have a forward connection to neuron $j$ as $\mathrm{I}_j$. The average number $B$ of a neuron's active connections then follows as

$$B = \frac{1}{N} \sum_{j=1}^{N} \sum_{i \in \mathbb{I}_j} q(T, t_{ij})$$

$$= \frac{C}{t_{max} - t_{min}} \int_{t_{min}}^{t_{max}} q(T, t) dt. \tag{S5}$$

When $t_{min} < T - \tau$ and $T + \tau < t_{max}$ the integral in the above equation equals approximately $2\tau$ and we obtain

$$B = \frac{2\tau C}{t_{max} - t_{min}}. \tag{S6}$$

In conjunction with Equation S4, this relation yields an analytical dependence of the fraction $a$ of active nodes on the connectivity $C$. This dependence is shown as a black line in Figure 2b and agrees excellently with numerical results. We find that $a = 0.5$ for a connectivity of $C_* \approx 1.85$.

**Analytical computation of pattern distances**

As described in the Introduction, we quantify the distance between two network activity patterns $\boldsymbol{x} = \left( x^{(1)}, x^{(2)}, ..., x^{(N)} \right)$ and $\boldsymbol{y} = \left( y^{(1)}, y^{(2)}, ..., y^{(N)} \right)$ through the relative Hamming distance

$$d(\boldsymbol{x}, \boldsymbol{y}) = \frac{1}{N} \sum_{i=1}^{N} \left| x^{(i)} - y^{(i)} \right|. \tag{S7}$$

Distinct patterns of network activity result from differences in the active connections. Denote by $\Delta B(\boldsymbol{x}, \boldsymbol{y})$ the average number of a neuron's active connections that differ between the two patterns $\boldsymbol{x}$ and $\boldsymbol{y}$. If this difference is small, as it is for the small differences in the signal period that we consider, the pattern distance $d(\boldsymbol{x}, \boldsymbol{y})$ can be approximated as depending linearly on the difference $\Delta B(\boldsymbol{x}, \boldsymbol{y})$

$$d(\boldsymbol{x}, \boldsymbol{y}) = \frac{da}{dB} \bigg|_{B_*} B(\boldsymbol{x}, \boldsymbol{y}) \tag{S8}$$

in which $B_*$ follows from the network connectivity $C_*$ through Equation S6.

We start by computing the distance $d(\boldsymbol{X}_T, \boldsymbol{X}_{T'})$ between the mean pattern $\boldsymbol{X}_T$ at period $T$ and the mean pattern $\boldsymbol{X}_{T'}$ at period $T'$. Consider a connection from a neuron $i$ to another neuron $j$ with a time delay $t_{ij}$. For the mean patterns we can ignore the phase noise in Equation S1: the connection is thus active during a signal period $T$ when $T\text{-}\tau < t_{ij} < T\text{+}\tau$ and vanishes otherwise. Analogously a signal period $T'$ yields an active connection when $T'\text{-}\tau < t_{ij} < T'\text{+}\tau$ and an inactive connection otherwise. Assume that $T<T'$; the other case follows by analogy. Only when the time delay $t_{ij}$ lies in the intervals $(T\text{-}\tau, T'\text{-}\tau)$ or $(T\text{+}\tau, T'\text{+}\tau)$ does the connection differ in its activity between the two patterns, and the probability of having such a time delay is $2|T'-T|/(t_{max} - t_{min})$. The average number $\Delta B(X_T, X_{T'})$ of a neuron's active connections that differ between the two patterns follows as

$$\Delta B(X_T, X_{T'}) = \frac{2C_*}{t_{max} - t_{min}} |T'-T| \tag{S9}$$

and the pattern distance reads

$$d(X_T, X_{T'}) = \frac{da}{dB}\bigg|_{B_*} \frac{2C_*}{t_{max} - t_{min}} |T'-T|. \tag{S10}$$

As observed numerically, the pattern distance increases linearly in $|T'\text{-}T|$ (Figure 3a). For the parameters employed in our simulations, the above expression yields a slope that is comparable but about 20% greater than the numerical value.

Let us now compute the distance $d(x_T, X_{T'})$ between a single-trial pattern $x_T$ during period $T$ and the mean pattern $X_{T'}$ during another period $T'$. This distance varies from trial to trial (Figure 3b). Again, we consider a connection from a neuron $i$ to another neuron $j$ that induces a certain time delay $t_{ij}$. As before, the mean pattern $X_{T'}$ is unaffected by the phase noise: the connection is active when $T'\text{-}\tau < t_{ij} < T'\text{+}\tau$ and vanishes otherwise. For the single-trial pattern $x_T$, however, the activity may fluctuate: as computed above, the connection is active with probability $q(T, t_{ij})$ and zero otherwise (Equation S4). To capture this stochasticity we introduce a random binary variable $Z_{ij}$ that is one when the connection from neuron $i$ to neuron $j$ differs in its activity between the single-trial pattern $x_T$ and the mean pattern $X_{T'}$; otherwise $Z_{ij}$ is zero. We find that $Z_{ij}$=1 with a probability $r(T, T', t_{ij})$ and $Z_{ij}$=0 with a probability $1\text{-}r(T, T', t_{ij})$ in which

$$r(T, T', t_{ij}) = \begin{cases} q(T, t_{ij}) & \text{if} \quad t_{ij} < T'\text{-}\tau \text{ or } t_{ij} > T'\text{+}\tau \\ 1 - q(T, t_{ij}) & \text{if} \quad T'\text{-}\tau < t_{ij} < T'\text{+}\tau \end{cases}. \tag{S11}$$

Because the difference $\Delta B(\boldsymbol{x}_T, \boldsymbol{X}_{T'})$ follows as the sum over many independent random variables $Z_{ij}$,

$$\Delta B(\boldsymbol{x}_T, \boldsymbol{X}_{T'}) = \frac{1}{N} \sum_{j=1}^{N} \sum_{i \in \mathbf{I}_j} Z_{ij}, \tag{S12}$$

the central limit theorem guarantees that, in the limit of a large system size $N$, the distribution of $\Delta B(\boldsymbol{x}_T, \boldsymbol{X}_{T'})$, and hence of the pattern distance $d(\boldsymbol{x}_T, \boldsymbol{X}_{T'})$ approaches a Gaussian distribution (Figure 3b). The mean value for the distribution of $\Delta B(\boldsymbol{x}_T, \boldsymbol{X}_{T'})$ reads

$$\left\langle \Delta B(\boldsymbol{x}_T, \boldsymbol{X}_{T'}) \right\rangle = \frac{1}{N} \sum_{j=1}^{N} \sum_{i \in \mathbf{I}_j} r(T, T', t_{ij}) \tag{S13}$$

and its variance is

$$\left\langle \Delta B(\boldsymbol{x}_T, \boldsymbol{X}_{T'})^2 - \left\langle \Delta B(\boldsymbol{x}_T, \boldsymbol{X}_{T'}) \right\rangle^2 \right\rangle = \frac{1}{N^2} \sum_{j=1}^{N} \sum_{i \in \mathbf{I}_j} r(T, T', t_{ij}) \Big[ 1 - r(T, T', t_{ij}) \Big]$$

$$= \frac{1}{N^2} \sum_{j=1}^{N} \sum_{i \in \mathbf{I}_j} q(T, t_{ij}) \Big[ 1 - q(T, t_{ij}) \Big]. \tag{S14}$$

The mean pattern distance follows from Equation S8 as

$$D(\boldsymbol{x}_T, \boldsymbol{X}_{T'}) = \frac{da}{dB}\bigg|_{B_*} \frac{1}{N} \sum_{j=1}^{N} \sum_{i \in \mathbf{I}_j} r(T, T', t_{ij}) \tag{S15}$$

and its variance $\sigma^2$ as

$$\sigma^2 = \left( \frac{da}{dB}\bigg|_{B_*} \right)^2 \frac{1}{N^2} \sum_{j=1}^{N} \sum_{i \in \mathbf{I}_j} q(T, t_{ij}) \Big[ 1 - q(T, t_{ij}) \Big]. \tag{S16}$$

The last equality shows that the variance does not depend on $T'$. We show that it is also independent of $T$ as long as $T$ is slightly greater than $t_{\min} + \tau$ and less than $t_{\max} + \tau$.

Because the delay $t_{ij}$ is chosen randomly from the interval $[t_{\min}, t_{\max}]$ the sums in Equations (S15) and (S16) can be expressed through integrals over time delays $t$ between $t_{\min}$ and $t_{\max}$:

$$D(\boldsymbol{x}_T, \boldsymbol{X}_{T'}) = \frac{da}{dB}\bigg|_{B_*} \frac{C_*}{t_{\max} - t_{\min}} \int_{t_{\min}}^{t_{\max}} r(T, T', t) dt \tag{S17}$$

and

$$\sigma^2 = \left(\frac{da}{dB}\bigg|_{B_*}\right)^2 \frac{C_*}{N(t_{max} - t_{min})} \int_{t_{min}}^{t_{max}} q(T,t)\big[1 - q(T,t)\big]dt.$$ (S18)

To compute the integral in Equation S18 we employ a piecewise linear approximation $q_{lin}(T,t)$ of $q(T,t)$ (Figure S2):

$$q_{lin}(T,t) = \begin{cases} 0 & \text{if} \quad t < T - \tau - \pi s/\sqrt{2L} \ \text{ or } \ t > T + \tau + \pi s/\sqrt{2L} \\ \frac{1}{2} + \frac{\sqrt{L}}{\sqrt{2}\pi s}(t - T + \tau) & \text{if} \quad T - \tau - \pi s/\sqrt{2L} < t < T - \tau + \pi s/\sqrt{2L} \\ 1 & \text{if} \quad T - \tau + \pi s/\sqrt{2L} < t < T + \tau - \pi s/\sqrt{2L} \\ \frac{1}{2} - \frac{\sqrt{L}}{\sqrt{2}\pi s}(t - T + \tau) & \text{if} \quad T + \tau - \pi s/\sqrt{2L} < t < T + \tau + \pi s/\sqrt{2L} \end{cases}.$$ (S19)

The integral then follows as

$$\int_{t_{min}}^{t_{max}} q_{lin}(T,t)\big[1 - q_{lin}(T,t)\big]dt = \frac{\sqrt{2}\pi s}{3\sqrt{L}}$$ (S20)

and we obtain

$$\sigma^2 = \left(\frac{da}{dB}\bigg|_{B_*}\right)^2 \frac{\sqrt{2}\pi s C_*}{3\sqrt{L}N(t_{max} - t_{min})}.$$ (S21)

The standard deviation $\sigma$ therefore decreases as $N^{-1/2}$ with an increasing system size $N$ and as $L^{-1/4}$ for a greater signal length $L$, in agreement with our numerical results (Figure 3c and 5a). The standard deviation is moreover independent of the signal periods $T$ and $T'$ as we have also found numerically (Figure 3b and 4). The values predicted by the above analytical expression are about 30% lower than those calculated numerically.

The linear approximation $q_{lin}(T,t)$ results, through Equation S11, in a piecewise linear approximation $r_{lin}(T,T',t)$ for $r(T,T',t)$ that we employ to approximate the integral in Equation S21 (Figure S2). Two cases then emerge. First, for small period differences $|T'-T| \leq \pi s/\sqrt{2L}$ we obtain

$$\int_{t_{min}}^{t_{max}} r_{lin}(T,T',t)dt = \frac{\pi^2 s^2 + 2L(T'-T)^2}{\sqrt{2L}\pi s}$$ (S22)

and a mean pattern distance of

$$D(\mathbf{x}_T, \mathbf{X}_{T'}) = \frac{da}{dB}\bigg|_{B_*} C_* \frac{\pi^2 s^2 + 2L(T'-T)^2}{\sqrt{2L}\pi s(t_{max} - t_{min})}.$$ (S23)

The pattern distance is quadratic in $T'$-$T$ for these small period differences, and has a nonvanishing minimum at $T'=T$ as seen in numerics (Figure 4).

Second, for larger period differences $|T'-T| \geq \pi s/\sqrt{2L}$ we compute

$$\int_{t_{\min}}^{t_{\max}} r_{\mathrm{lin}}(T,T',t)dt = 2|T'-T| \qquad (S24)$$

and obtain

$$D(\boldsymbol{x}_T,\boldsymbol{X}_{T'}) = \frac{da}{dB}\bigg|_{B_*} \frac{2C_*}{t_{\max}-t_{\min}}|T'-T|. \qquad (S25)$$

For these larger period differences the pattern distance therefore increases linearly in $|T'-T|$ as we have already found numerically (Figure 4). Because the value $\Delta T = \pi s/\sqrt{2L}$ separates the two regimes, we consider it to be the network's threshold for period discrimination. The above analytical expression shows that $\Delta T$ is independent of the system size $N$ but decreases according to $L^{-1/2}$ with increasing signal length $L$, in agreement with our numerical results (Figure 5b).

Equations S23 and S25 show that the mean distance $D(\boldsymbol{x}_T,\boldsymbol{X}_{T'})$ is invariant under exchange of $T$ and $T'$: $D(\boldsymbol{x}_T,\boldsymbol{X}_{T'})=D(\boldsymbol{x}_{T'},\boldsymbol{X}_T)$ in agreement with numerical results. Because the standard deviation $\sigma$ does not dependent on either $T$ or $T'$, it follows that the distributions of $d(\boldsymbol{x}_T,\boldsymbol{X}_{T'})$ and $d(\boldsymbol{x}_{T'},\boldsymbol{X}_T)$ are identical.

**Supplementary Figure 1: Schematic illustrations of phase locking.** A stimulus at a frequency below 800 Hz elicits a series of action potentials in, for example, spherical bushy cells of the auditory brain stem; spikes occur at every cycle of stimulation and at a preferred phase. Because an action potential lasts about 1 ms, those neurons no longer fire spikes at every cycle when the signal frequency exceeds 800 Hz. Spikes still occur, however, at a preferred phase. Above 4 kHz the spikes' phases are random.

**Supplementary Figure 2: The probabilities $q(T,t)$ and $r(T,T',t)$ and their linear approximations. a**, The probability $q(T,t)$ (red) is approximately one for $t$-$T$ between -$\tau$ and $\tau$ and zero outside this interval. The transitions at -$\tau$ and $\tau$ (grey lines) sharpen for greater signal length; here we have employed $L$=10. The piecewise linear approximation $q_{\text{lin}}(T,t)$, Equation 19, is shown as black dashed line. **b**, The probability $r(T,T',t)$ (red) and its piecewise-linear approximation $r_{\text{lin}}(T,T',t)$ (black dashed) are nonzero in only a small interval of $t$-$T$ around -$\tau$ and $\tau$ (gray lines). The curves have been obtained from $L$=10 and $T'$=1.02 $T$.

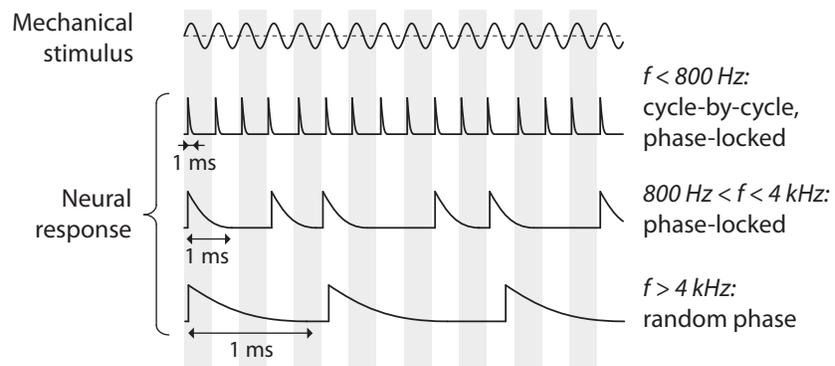

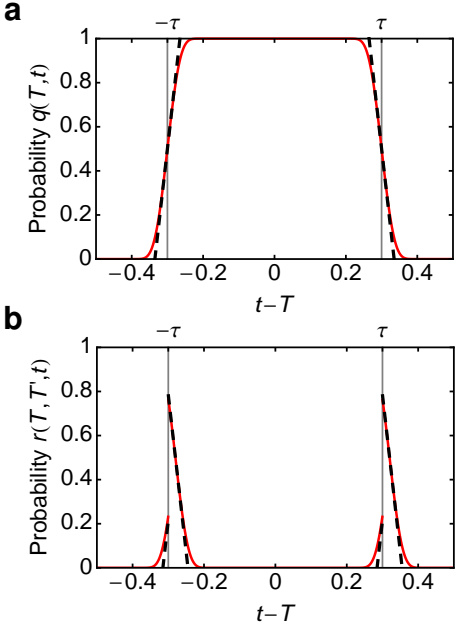